\documentclass{fullpaper_hutech}

\begin{document}

\title{Mechanical Properties of Wilberforce Pendulum}
\twocolumn[
\begin{@twocolumnfalse}
\begin{flushleft}
\maketitle
\begin{abstract}
\textbf{Sanghwa Lee}\\
\vspace{0.5cm}
torytony24@gmail.com\\
\vspace{0.5cm}
Korea Science Academy of KAIST, 20-078\\
\vspace{0.5cm}
\textbf{ABSTRACT}\\
This paper shows the study of interesting mechanical properties of Wilberforce pendulum. Analyzing qualitatively of the pendulum, it is able to know how the phenomenon occurs. By setting of the quantitative model, equation of the motion is derived. Considering the mass and moment of inertia of the spring, the experiment was done by changing the moment of inertia as the main parameter. The results were analyzed by defining oscillation ratio and conversion factor. 
\end{abstract}
\vspace{20pt}
\end{flushleft}
\end{@twocolumnfalse}
]
\thispagestyle{firstpage}

\lstset{
  basicstyle=\ttfamily,
}

\section{INTRODUCTION}

This study is about the mechanical properties of the Wilberforce pendulum. A Wilberforce pendulum consists of a mass hanging from a vertically oriented helical spring. The mass can both move up and down on the spring and rotate about its vertical axis. The theoretical analysis, which consists of qualitative and quantitative analysis will be done through mechanical investigation of the system.

Experiments will be done using the mechanical system, which is called Wilberforce pendulum, and analysis of the pendulum will be done using various programs. Investigating the data that is earned through experiments, the behavior of the pendulum is investigated by comparing both theoretical results and experimental results. In addition, the motion of Wilberforce pendulum depends on relevant parameters, so that analysis based on considering relevant parameters will be done.

\section{THEORETICAL MODEL}

A theoretical model was set up based on the qualitative analysis of the pendulum. Based on the qualitative analysis and the phenomenon observation, the theoretical model was set up in two different ways.

\subsection{Qualitative anlysis}

There are two perspectives on the spring. From a perspective of force, the increase in the vertical displacement leads to the rotational increase of the angular displacement. Vice versa, the increase of the angular displacement leads to an increase in vertical displacement. The restoring force acts on the spring, both vertically and torsionally, which makes sense to the decrease in vertical displacement and rotational displacement. As a perspective in energy, periodic transfer between rotational energy and translational energy occurs on the pendulum as it shows the periodic motion.

In terms of phenomenon observation, beating occurs in the motion of the pendulum. If the angular frequency of vertical motion and rotational motion is denoted $\omega _z$ and $\omega _\theta $ respectively, the beating phenomena can be explained. When $\omega _z > \omega _\theta $ satisfies, the energy conversion is unclear. It is able to observe that the beating occurs through two motions since the energy is not perfectly transferred. In the same way, when $\omega _z < \omega _\theta $ satisfies, the energy conversion is unclear that the beating occurs. However, for the singularity point that $\omega _z = \omega _\theta $, the energy conversion is significant when observed. The resonance occurs between two oscillations that a large amplitude is observed. As we define oscillation ratio $\Omega=\omega_z / \omega _\theta $, the more $\Omega$ is closer to 1, the resonance is more clearly shown. The resonance occurs most clearly when $\Omega=1$.

\subsection{Reason of phenomenon}

As analyzed in qualitative analysis, the restoring force acts on the spring so that two motions act on each other. This was explained quantitatively by Clive L. Dym \cite{lambda}. By Castigliano's Second Theorem, the complementary energy is denoted as the following equation.
\begin{equation}
\begin{split}
U^{*} =& \int_{0}^{L_c} \left[ \frac{M_x^2}{2GJ} + \frac{M_{z^*}^2}{2EI} + \frac{V^2}{2GA} + \frac{N^2}{2EA} \right]dx\\
=& \frac{1}{2}f_{11}F^2 - 2f_{12}FT + \frac{1}{2} f_{22}T^2
\end{split}
\end{equation}
The first term consisting $f_{11}$ shows the elastic potential energy, while the last term consisting $f_{22}$ shows the torsional potential energy. The second term consisting $f_{12}$ shows the coupling energy. This coupling energy is the source of the beating. It is shown that beating occurs from the perspective of energy.

\subsection{Quantitative model}

Setting up the coordinate system as a cylindrical coordinate system would be easier to analyze. The system can be set as \textbf{Fig. 1}.

\begin{figure}
\begin{center}
\includegraphics[width=6cm]{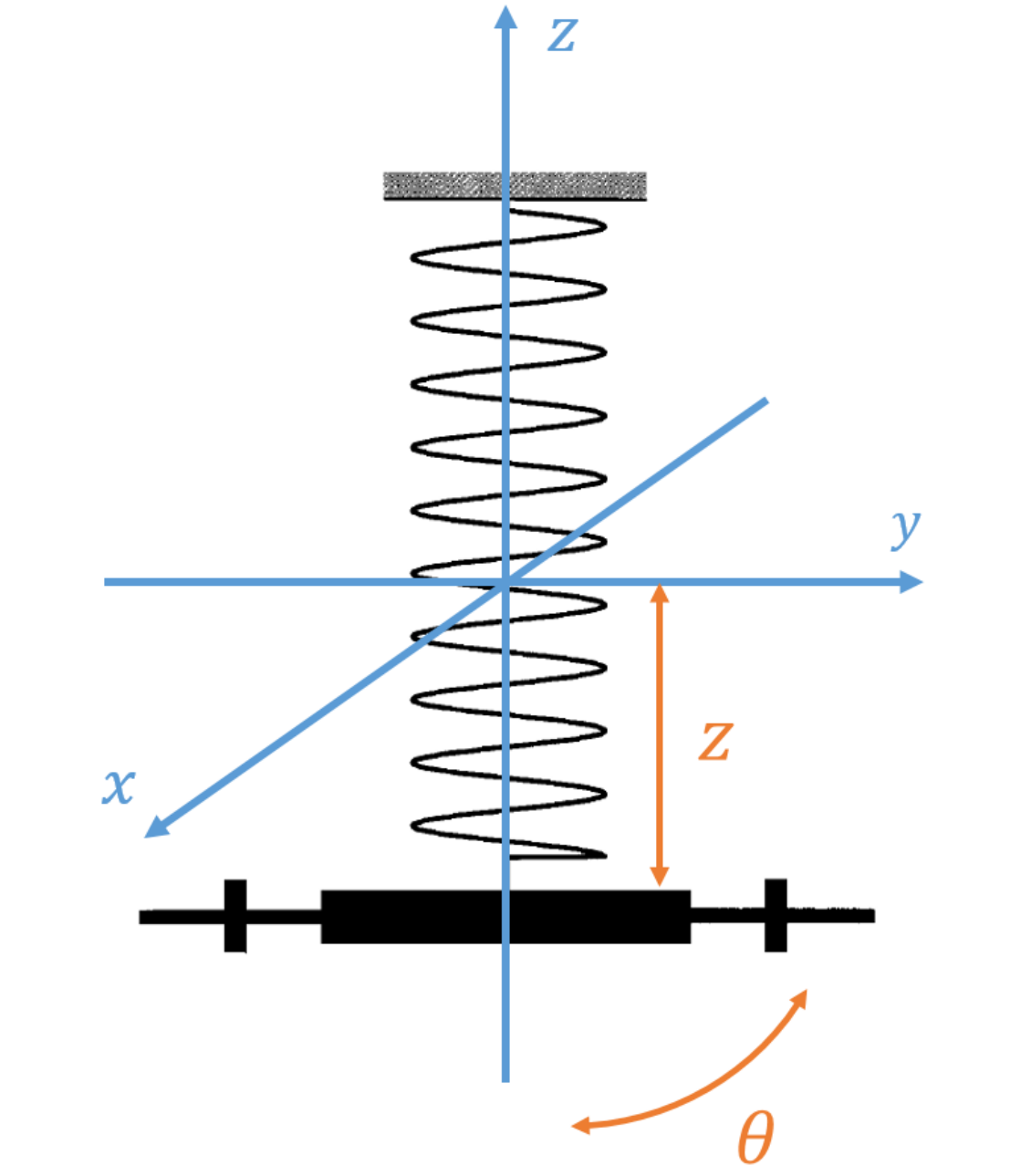}
\caption{Coordinate system}
\label{figure::label}
\end{center}
\end{figure}

Set vertical displacement as $z$, angular displacement as $\theta$, vertical spring constant as $k$, torsional spring constant as $\kappa$, the mass of the bob as $m$, and the moment of inertia of the bob as $I$.

As the reason of the phenomenon is discussed at \textbf{2.2}, considering the coupling energy term, it is able to predict that coupling spring constant exists other than vertical spring constant and torsional spring constant. Taking Taylor expansion of potential energy of the spring,
\begin{equation}
\begin{split}
&V(z,\theta)\\
&= V(z_0 , \theta_0 ) + \left( z \frac{\partial V(z_0, \theta_0)}{\partial z} + \theta \frac{\partial V(z_0, \theta _0 )}{\partial \theta} \right)+\\
&\frac{1}{2!} \left( z^2 \frac{\partial ^2 V(z_0, \theta_0)}{\partial z^2} + 2z\theta \frac{\partial ^2 V(z_0 , \theta _0 ) }{\partial z \partial \theta }+ \theta^2 \frac{\partial ^2 V (z_0 , \theta _0 ) }{\partial \theta ^2} \right)\\
&+ \cdots
\end{split}
\end{equation}
There exist terms $1/2kz^2$ and $1/2 \kappa \theta ^2$, and there is one more term other than those two. The second-power approximation shows the coupling term $\lambda z \theta$, where $\lambda$ is constant.

\subsection{Equation of motion}

Based on the quantitative model, the force acting on the pendulum can be interpreted through energy. Using the force-potential relation $F= -\nabla U$,
\begin{equation}
\begin{split}
m \ddot{z} + kz + \frac{\partial (\lambda z \theta)}{\partial z} =& m \ddot{z} + kz + \lambda \theta =0\\
I \ddot{\theta} + \kappa \theta + \frac{\partial (\lambda z \theta )}{\partial \theta} \rho =& I \ddot{\theta} + \kappa \theta + \lambda z =0
\end{split}
\end{equation}
Initial conditions are set as the following,
\begin{equation}
\begin{split}
z(0) =& z_0\\
\dot{z}(0) = \theta(0) =& \dot{\theta}(0) =0
\end{split}
\end{equation}
It is able to substitude the angular frequency as the following.
\begin{equation}
\begin{split}
\omega_z =& \sqrt{\frac{k}{m}}\\
\omega_\theta =& \sqrt{\frac{\kappa}{I}}
\end{split}
\end{equation}
Solving the general solution,
\begin{equation}
\begin{split}
z(t)=&\frac{z_0}{\omega_1^2 - \omega_2 ^2} \left( (\omega_z^2 - \omega_z^2) \cos{\omega_1 t} - (\omega_z^2 - \omega_1 ^2 ) \cos{\omega_2 t} \right)\\
\theta(t) =& \frac{-z_0 (\omega_1^2 -\omega_z^2 ) ( \omega_2^2 - \omega_z^2 ) }{\lambda( \omega_1^2 - \omega_2^2 )} (\cos{\omega_1 t} - \cos{\omega_2 t})
\end{split}
\end{equation}
where
\begin{equation}
\begin{split}
\omega_{1,2} = \sqrt{ \frac{ (\omega_z^2 + \omega_{\theta}^2) \pm \sqrt{ (\omega_z^2 + \omega_{\theta}^2)^2 - 4 \left( \omega_z^2 \omega_{\theta}^2 - \frac{\lambda^2 }{mI} \right) } }{2} }
\end{split}
\end{equation}

\section{EXPERIMENT}

\subsection{Experimental setup \& measurement}

The 3D-Model of the pendulum was first made. The pendulum has a slinky connected to the bob, and the rod intersects the bob. Two weights are connected at a distance $d$ from the bob, symmetrically. Based on the model, the pendulum was made with slinky, rod, and weight. The m

The apparatus of the experiment includes the pendulum with slinky connected to the ceiling and PASCO Capstone Motion Sensor on the floor. The distance between the pendulum and sensor was measured 0.62 m. For the rotational motion of the bob, data was extracted through video analysis using TRACKER.

Measuring the mass and the length properties of the bob, it is able to know the values $m_b$ and $I_b$. To be more accurate, the mass and moment of inertia of the spring should also be considered. When a spring is enlonged by $x$, consider a piece of spring at distance $\xi$. Then using the following relationship,
\begin{equation}
\begin{split}
\dot{\xi} = \frac{\xi}{x} \dot{x}
\end{split}
\end{equation}
Considering kinetic energy of the spring,
\begin{equation}
\begin{split}
&E_K = \int_0^x \frac{1}{2} \dot{x}^2 dm\\
&= \int_0^x \frac{1}{2} \frac{d\xi}{x}m\frac{\xi^2}{x^2}\dot{x}^2 = \frac{1}{2} \left( \frac{1}{3} m \right) \dot{x}^2
\end{split}
\end{equation}
Then the effective mass of the whole system $m_{eff}$ can be interpreted with the mass of the bob $m_b$ and the mass of the spring $m_s$.
\begin{equation}
\begin{split}
m_{eff}=m_b +\frac{1}{3} m_s
\end{split}
\end{equation}
In same way, the effective moment of inertia of the whole system $I_{eff}$ can be interpreted with moment of inertia of the bob $I_b$ and the mass of the spring $m_s$.
\begin{equation}
\begin{split}
I_{eff}=I_b +\frac{1}{3} m_s R^2
\end{split}
\end{equation}

\subsection{Experimental results}

As shown in \textbf{Fig. 2} down below, it is able to see the resonance between the vertical motion (dark blue) and the rotational motion (light blue) clearly.

\begin{figure}
\begin{center}
\includegraphics[width=8cm]{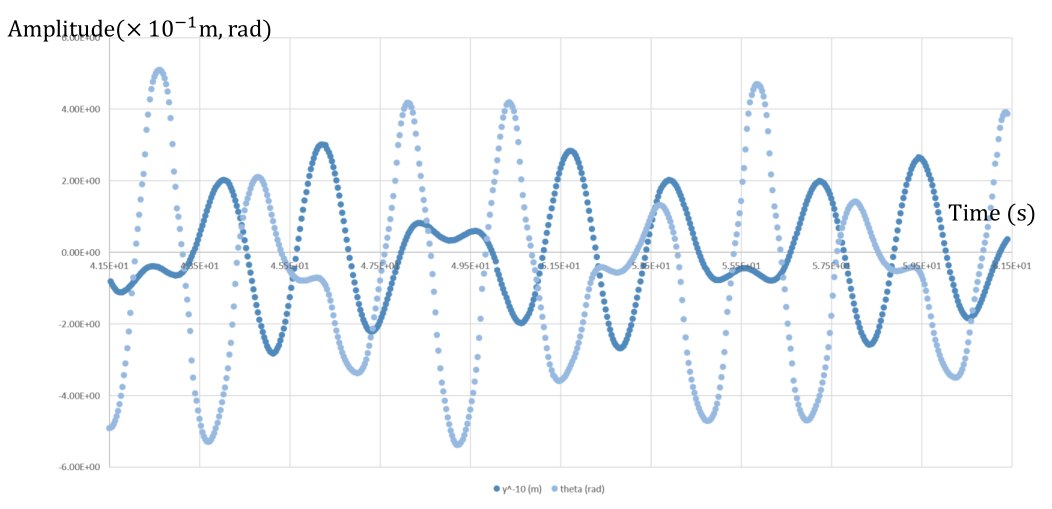}
\caption{Resonance graph}
\label{figure::label}
\end{center}
\end{figure}

The distance between the center of the bob and weight, which is denoted as $d$, is varied from 2 cm to 9 cm. Fourier analysis was done for the vertical data through MATLAB. The MATLAB code for FFT is shown in \textbf{Appendix 1}. By analyzing two modes of the pendulum through Fourier analysis, the spring constant was calculated by considering it as a fitting parameter. As shown in \textbf{Fig. 3}, FFT data clearly shows two peaks, which means that the oscillation is a combination of two different modes beating.

\begin{figure}
\begin{center}
\includegraphics[width=6cm]{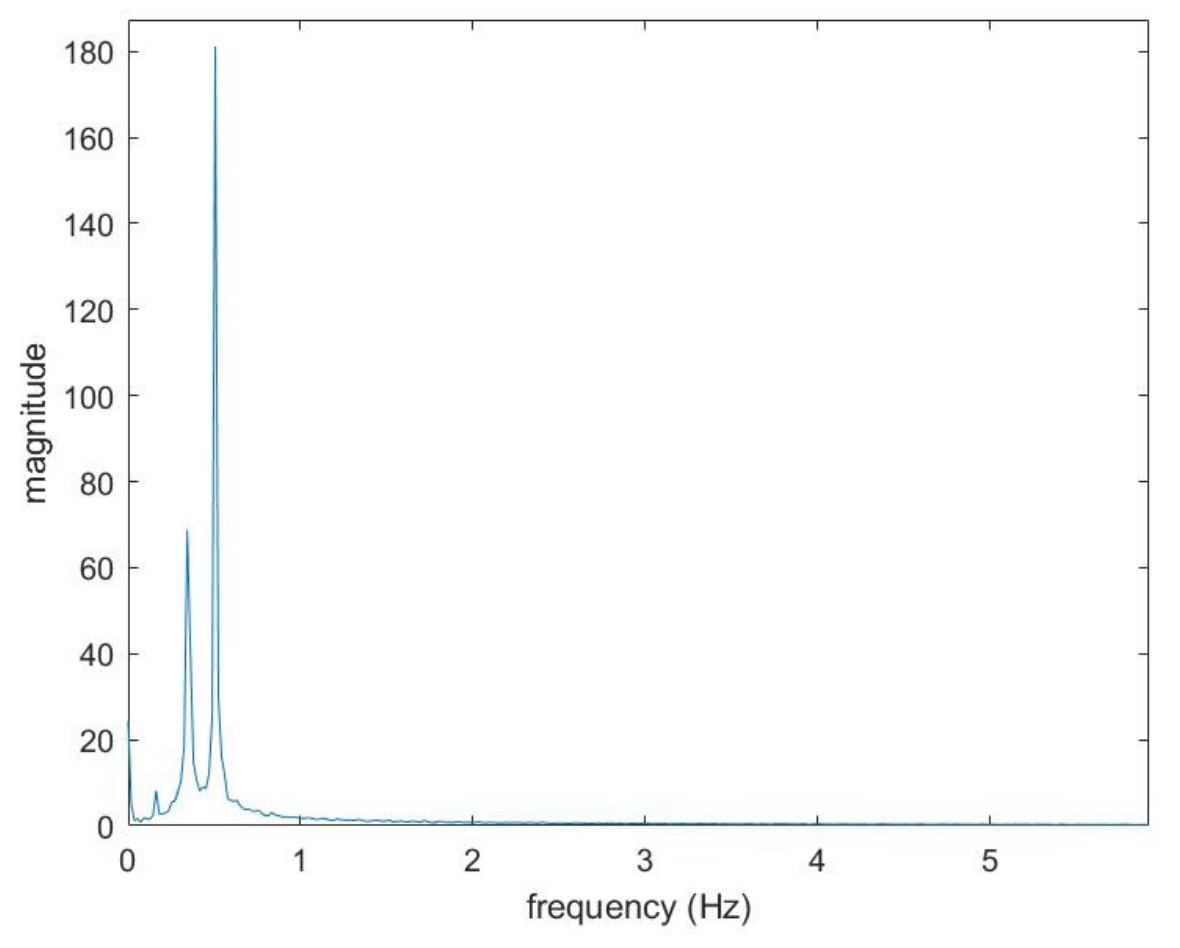}
\caption{Fast Fourier Transform}
\label{figure::label}
\end{center}
\end{figure}

Taking the vertical angular frequency as a fitting parameter, the theory and experiment comparison is shown in the table below. The absolute value of the error rate between the theoretical vertical angular frequency and the experimental angular frequency is lower than 4\%, which means that the experiment and theory match well. Other experimental graphs are shown in \textbf{Appendix 2}.

\begin{center}
\begin{tabular}{ |c|c|c|c|c|c| }
\hline
$d$ (cm) & 2.2 & 4.4 & 5.2 & 6.1 & 8.9 \\
\hline\hline
$\omega_z$ (rad/s) & 2.82 & 2.90 & 2.82 & 2.89 & 2.89 \\
\hline
Error rate (\%) & -0.71 & -3.57 & -0.71 & -3.21 & -3.21\\
\hline
\end{tabular}
\end{center}

\begin{center}
\textbf{Table. 1.} Experimental results table
\end{center}

\section{ANALYSIS}

\subsection{Regression}

Since we have varied the distance between the middle of the spring and the displacement of the bob, we have changed the moment of inertia through the experiment. It is able to plot the angular frequency of each mode through the change of the moment of inertia. The following graph was drawn by fitting the other spring constants, rotational spring constant, and coupling spring constants. These were set as fitting parameters.

\begin{figure}[h]
\begin{center}
\includegraphics[width=9cm]{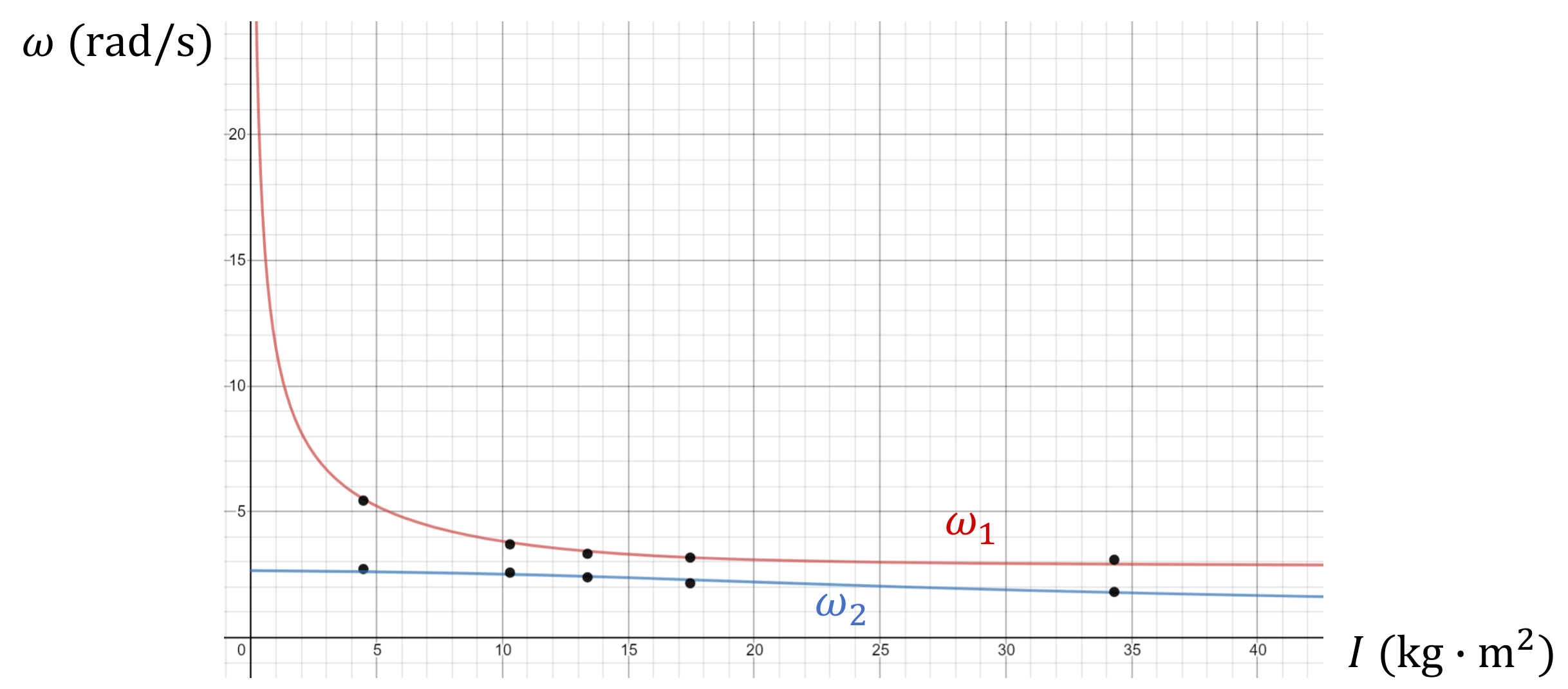}
\caption{Regression graph}
\label{figure::label}
\end{center}
\end{figure}

As the regression graph is drawn with DESMOS, which is the graph between the moment of inertia and angular frequency of the pendulum. The angular frequency $\omega_1$ and $\omega_2$ is graphed as \textbf{Fig. 4}. The regression formula is shown in equation (6). The regression formula is an equation that was derived in theoretical analysis, and it fits the experimental result well. This clearly shows that the theoretical result and the experimental result matches well.

\subsection{Oscillation ratio}

Oscillation ration $\Omega$ was set up to explain the observation of the phenomenon, on the perspective of energy transfer between translational energy and rotational energy, which is denoted $\Omega=\omega_z / \omega _\theta$. It was also observable that as the oscillation ratio gets closer to 1, the energy transfer gets closer to 100 percent. It is able to analyze that information quantitatively. Defining the beat angular frequency $\omega_b$, which is introduced as $\omega_b=\omega_1-\omega_2$, It is able to find the relationship between $\Omega$. As the main goal is to compare the difference with $\Omega$ and 1, defining $\Gamma $ as the following will help the explanation.

\begin{equation}
\begin{split}
\Gamma = |\Omega-1| = \left|\frac{\omega_z}{\omega_\theta}-1 \right|
\end{split}
\end{equation}

As a heuristic, it is able to predict that $\Gamma$ increases as $\omega_b$ increases, and $\Gamma$ decreases as $\omega_b$ decreases. Introducing experimental data as \textbf{Table. 2},

\begin{center}
\begin{tabular}{ |c|c|c| }
\hline
$d$ (cm) & $\Gamma$ & $\omega_b$ (rad/s) \\
\hline\hline
2.2 & 0.4766 & 2.720\\
\hline
4.4 & 0.1601 & 1.125\\
\hline
5.2 & 0.0507 & 0.9354\\
\hline
6.1 & 0.146 & 1.020\\
\hline
8.9 & 0.370 & 1.277\\
\hline
\end{tabular}
\end{center}

\begin{center}
\textbf{Table. 2.} Oscillation ratio related to beat
\end{center}

\subsection{Conversion factor}

In order to focus on the idea of the energy conversion between translational energy and rotational energy, it is able to define a conversion factor $C_E$. The conversion factor of energy is defined similarly to the Quality factor of forced oscillation, which is denoted

\begin{equation}
\begin{split}
C_E=\frac{\Delta E}{E_{max}}
\end{split}
\end{equation}

It shows as \textbf{Table. 3} as implementing experimental data.

\begin{center}
\begin{tabular}{ |c|c|c| }
\hline
$d$ (cm) & $\Delta E$ & $C_E$ \\
\hline\hline
2.2 & 0.065 & 0.255\\
\hline
4.4 & 0.171 & 0.074\\
\hline
5.2 & 16.7 & 0.988\\
\hline
6.1 & 0.152 & 0.844\\
\hline
8.9 & 0.076 & 0.295\\
\hline
\end{tabular}
\end{center}

\begin{center}
\textbf{Table. 3.} Conversion factor experimental data
\end{center}

It is clearly shown in an experimental method that the conversion factor is close to 1 as the motion is close to resonance. This way, the relationship between the efficiency of energy transfer and beat is explained experimentally.

\section{CONCLUSION}

The study of mechanical properties of Wilberforce pendulum was done by theoretical analysis with experimental analysis. The theoretical model was set up based on the observation of the phenomenon with qualitative analysis. Analyzing in two different perspectives, force, and energy, it is able to explain the phenomenon. Especially for the energy perspective, some parameters were defined and introduced for a quantitative explanation using experimental data. The reason for the phenomenon was explained using the research of Clive L. Dym\cite{lambda}, which used Castigliano's Second Theorem. A quantitative model was set by designing the coordinate system. The existence of the coupling term was shown by Taylor expansion of potential energy, and by using it, a mechanical equation of motion was derived. By solving the differential equation, the general solution was derived.

Experimental setup and measurement were done by firstly designing the 3D-Model of the pendulum. The experiment was done several times with PASCO Capstone, and data was extracted using TRACKER. Using the kinetic energy integration, mass and moment of inertia of spring were considered as well as the bob. Experimental results are shown by analyzing experimental data using Fast Fourier Transform (FFT) and showing the error rate in the source of the table. Analysis of the experiment was done in three ways, including drawing a regression graph using DESMOS. The oscillation ratio, which was defined in the qualitative analysis, was reorganized by defining $\Gamma$, and implementing experimental data. Also, conversion factor $C_E$ is defined, and experimental data was implemented for analysis.

\section*{Appendix 1}

Following is the MATLAB code for FFT.

\begin{lstlisting}
y = fft(x);
N = length(y);
Fs = 1/t;
f = (0 : N-1) * Fs / N;
plot(f , abs(y))
\end{lstlisting}

\end{document}